\documentstyle[prl,aps,amssymb,epsfig,multicol]{revtex}
\begin{document}
\draft

%

\title{Mid-Rapidity Neutral Pion Production in Proton-Proton Collisions 
at $\mathbf{\sqrt{s}}$=200~GeV}

%


\author{
S.S.~Adler,$^{5}$
S.~Afanasiev,$^{17}$
C.~Aidala,$^{5}$
N.N.~Ajitanand,$^{43}$
Y.~Akiba,$^{20,38}$
J.~Alexander,$^{43}$
R.~Amirikas,$^{12}$
L.~Aphecetche,$^{45}$
S.H.~Aronson,$^{5}$
R.~Averbeck,$^{44}$
% ...
\\(PHENIX Collaboration)\\
}
\address{
% ...
$^{5}$Brookhaven National Laboratory, Upton, NY 11973-5000, USA\\
% ...
$^{12}$Florida State University, Tallahassee, FL 32306, USA\\
% ...
$^{17}$Joint Institute for Nuclear Research, 141980 Dubna, Moscow Region, Russia\\
% ...
$^{20}$KEK, High Energy Accelerator Research Organization, Tsukuba-shi, Ibaraki-ken 305-0801, Japan\\
% ...
$^{38}$RIKEN (The Institute of Physical and Chemical Research), Wako, Saitama 351-0198, JAPAN\\
% ...
$^{43}$Chemistry Department, Stony Brook University, SUNY, Stony Brook, NY 11794-3400, USA\\
$^{44}$Department of Physics and Astronomy, Stony Brook University, SUNY, Stony Brook, NY 11794, USA\\
$^{45}$SUBATECH (Ecole des Mines de Nantes, CNRS-IN2P3, Universit{\'e} de Nantes) BP 20722 - 44307, Nantes, France\\
% ...
}

\date{\today}        
\maketitle

%

\begin{abstract}
The invariant differential cross section for inclusive neutral pion production in
$p\!+\!p$ collisions at $\sqrt{s}$~=~200~GeV has been measured at mid-rapidity
($|\eta|\!<\!0.35$) over the range $ 1\!<\!p_T\! \alt \!14$~GeV/$c$
by the PHENIX experiment at RHIC. 
Predictions of next-to-leading order perturbative QCD calculations
are consistent with these measurements. The precision of our result is sufficient 
to differentiate between prevailing gluon-to-pion fragmentation functions.
\end{abstract}
\pacs{PACS numbers: 13.85.Ni, 13.87.Fh, 25.75-q, 25.75.Dw}

\begin{multicols}{2}   
\narrowtext            
%
%
%
%
%
%
%
%
%

Particle production at large transverse momenta, $p_T$, in hadronic reactions
has provided an important testing ground for 
perturbative Quantum Chromodynamics (pQCD)~\cite{Owens}.
Next-to-leading order pQCD calculations describe Tevatron ($\sqrt{s}$$=$$1.8$~TeV) 
measurements of 
inclusive jet production~\cite{Tevatron_Jet} within 10\% and direct photon
production~\cite{Tevatron_Photon}, in 
which the elementary quark-gluon scattering produces a photon directly,
within 50\%.  
For high-$p_T$ pion production, the recent calculations have not 
been compared with the UA2 data~\cite{UA2} obtained at $\sqrt{s}$$=$540~GeV.
However, at lower center of mass energies ($\sqrt{s}$$\alt$63~GeV), they 
underestimate the data by a factor of $\sim$2.5~\cite{aur_pi0}.
Similar discrepancies have been observed for direct photon measurements from fixed
target experiments~\cite{vogelsang} and have been attributed to effects of soft-gluon 
radiations beyond NLO~\cite{nlo}, to effects of initial intrinsic transverse 
momentum, $k_T$~\cite{apan1}, or to experimental problems in the 
difficult direct photon measurements~\cite{aur_gamma,apan}. 
The $\pi^{\circ}$ calculations, as compared to the jet or direct photon
calculations, also require the probability for the scattered quark or gluon
to fragment into a pion.  

Information on fragmentation to pions~\cite{frag_theory:bkk,frag_theory:kkp,frag_theory:kretzer,frag_data,frag_theory:biebel} 
has principally come from global analyses of inclusive hadron 
production in $e^{+}e^{-}$ annihilation.  
These analyses constrain the quark-to-pion fragmentation
functions well but, via the scale dependence, the gluon-to-pion fragmentation function
to a lesser extent.
For example, the latter function at a scale of 100~GeV$^{2}$
can vary by a factor of 2 to 10 when the fraction of the
initial gluon momentum carried by the pion is above 0.5.
The more direct 
measurements of gluon fragmentation functions from $b$-tagged, three jet event data 
from LEP~\cite{LEP} have played a limited role
in the global analyses 
because NLO corrections are unavailable for the quantitative treatment, including
scale and scheme dependences, of these data.
As has been explored for measurements of inclusive hadron
production in $p$$+$$\bar{p}$ collisions~\cite{frag_theory:incl_hadrons},
results from inclusive pion production at 
high $p_T$ can be included in the global analyses and thus
may provide meaningful constraints on the gluon-to-pion fragmentation.
These results will also provide a reference needed for quantifying the
suppression of $\pi^{\circ}$ production observed in Au-Au collisions at
the Relativistic Heavy Ion Collider (RHIC)~\cite{ppg014} and, to the extent 
of agreement with the calculations, the foundations for the planned
polarized gluon density measurement with polarized protons in
RHIC~\cite{spin}. 

In this Letter, we report the first results on inclusive neutral pion 
production in $p\!+\!p$ collisions at a center of mass energy ($\sqrt{s}$) 
of 200~GeV as extracted 
from the data collected during the 2001-2002 run period (Run-2) of RHIC.
The bunched proton beams in the collider were vertically polarized with
spin orientations alternating in successive bunches.  By balancing the 
integrated luminosity
in the different spin states, the effects from polarization were canceled at 
the 0.1\% level.

In Run-2, the PHENIX experiment~\cite{phenix_nim} operated two central arm 
spectrometers, one muon arm spectrometer, and other detectors 
for triggering and vertex determination.  This work
used the beam-beam counters (BBC)~\cite{nim_bbc} for determining 
the collision vertex and constructing the 
minimum bias (MB) trigger,
and the electromagnetic calorimeters (EMCal)~\cite{nim_emc} for detecting the 
neutral pions and deriving high-$p_T$ triggers. 

The unbiased differential cross section for $\pi^{\circ}$ production is
calculated from the MB triggered data sample as
\begin{equation}
E \frac{d^3 \sigma}{dp^3} = \frac{1}{\hat{\cal L}}
                            \cdot
                            \frac{1}{2 \pi p_T^{*}} 
                            \cdot 
                            \frac{C_{\rm reco} \cdot C_{\rm conv}}{f_{\pi^{\circ}}}
                            \cdot
                            \frac{N_{\pi^{\circ}}}{\Delta p_T \Delta y},
\label{eq:mbcs}
\end{equation}
where 
$N_{\pi^{\circ}}$ is the number of $\pi^{\circ}$'s observed in a $\Delta p_T$ wide bin at 
$p_T^{*}$ defined as the $p_T$ for which the cross section equals its 
average over the bin;
$\Delta y$ is the rapidity range;
$C_{\rm reco}$ is a correction for the acceptance, reconstruction 
efficiency, and $p_T$ smearing;
$C_{\rm conv}$ is a correction for the conversion of decay photons;
$f_{\pi^{\circ}}$ is the fraction of the inclusive $\pi^{\circ}$ yield for which 
the MB trigger condition was satisfied; and
$\hat{\cal L}$ is the integrated luminosity for the analyzed data 
sample.
The high-$p_T$ triggered sample required an additional correction 
to account for the 
efficiency of this trigger for $\pi^{\circ}$ detection.

The MB trigger imposed the requirement that the collision vertex
was within 75~cm of the center of the interaction region.   This vertex
was reconstructed from the difference in the arrival times of 
particles at the BBCs which were located along the beam line 
at $\pm$1.44~m from the nominal
interaction point and subtended the pseudorapidity
range $\pm$(3.0-3.9) with full azimuthal coverage.  
In the analysis of the data, 
a more restrictive requirement of $\pm$30~cm was applied.

The EMCal consisted of two subsystems: 
a six sector, lead scintillator 
(PbSc) calorimeter and a two sector, lead glass (PbGl) calorimeter.
Located 
at a radial distance of $\sim$5~m from the beam line,
each of these sectors covered the pseudorapidity range of $|\eta|$$<$$0.35$ and
an azimuthal angle interval of $\Delta\phi \! \approx \! 22.5^{\circ}$.
Each of the towers in the calorimeter subtended 
$\Delta \phi \! \times \! \Delta \eta \sim 0.01 \! \times \! 0.01$, thus ensuring that
the two photons from a decayed $\pi^{\circ}$ were resolved up to 
a $p_T$ of at least 20~GeV/$c$.
The energy calibration was corroborated by the position of 
the $\pi^{\circ}$ invariant mass peak, the energy deposit from minimum 
ionizing charged particles traversing the EMCal (PbSc), and 
the correlation between the energy deposit in the EMCal and
the measured momentum for
electrons and positrons identified by the ring-imaging \v{C}erenkov 
detector.  
These studies showed that the accuracy of the energy measurement  
was within 1.5\%. 
At a $p_T$ of $\sim$11~GeV/$c$, this uncertainty translates into a systematic error on the 
$\pi^{\circ}$ yield of $\sim$12\%. 
The effective energy resolution for the dataset was deduced from the widths of the $\pi^{\circ}$ 
mass peaks, which varied with $p_T$ from 7\% to 10\% (PbSc) and 12\% to 13\% (PbGl),
and a comparison of the measured energy and momentum for the
identified electrons and positrons.  

\begin{figure}[ht]
\centerline{\epsfig{file=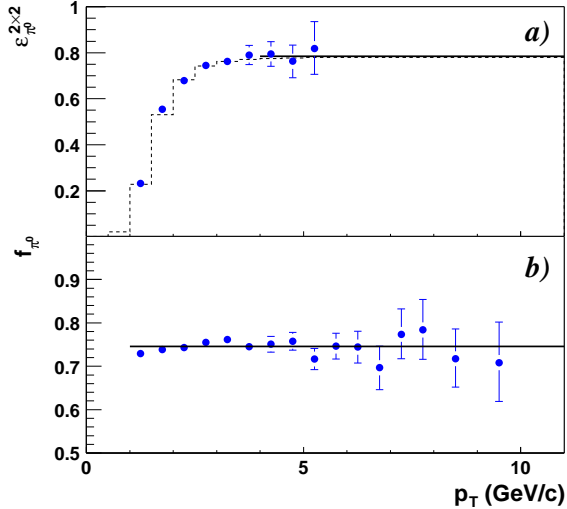,width=1.0\linewidth}}
\caption{a) The efficiency of the 2$\times$2 high-$p_T$ trigger 
            for $\pi^{\circ}$'s as a function of the $p_T$ of the $\pi^{\circ}$.   
            The dashed and solid lines show the results of 
            a Monte Carlo simulation based on the 2$\times$2 trigger 
            tile efficiencies and the limit derived
            from the fraction of active trigger tiles, respectively.
         b) The fraction of the inclusive $\pi^{\circ}$ yield
            which satisfied the MB trigger condition.
            The solid line shows a fit of these data to a constant.}
\label{fig:ERT} 
\end{figure}

The number of recorded high-$p_T$ $\pi^{\circ}$'s was enhanced
by a high-$p_T$ trigger (denoted as 2$\times$2) in which
threshold 
discrimination was applied independently to sums of the analog signals 
from non-overlapping, 2$\times$2 groupings (called tiles) of 
adjacent EMCal towers.
During this run, the thresholds corresponded to a 
deposited energy of 0.75~GeV.
The efficiency of this trigger for $\pi^{\circ}$ detection,
$\varepsilon_{\pi^{\circ}}^{\mathrm{2\times2}}(p_T)$, was obtained from the
MB data.  As shown in
Fig.~\ref{fig:ERT}a, 
this efficiency reached a plateau at a
$p_T$ of $\sim$3~GeV/$c$.  This dependence 
was reproduced by Monte Carlo calculations which included the
measured tile threshold curves, the EMCal detector response, and the
geometry of the active trigger tiles.
The saturation level, $0.78\pm0.03$ for both PbSc 
and PbGl, was consistent with the geometrical 
acceptance of the active trigger tiles.  For $p_T$$>$4~GeV/$c$, the
geometrical acceptance was used to correct the
$\pi^{\circ}$ yield in the high-$p_T$ sample.

Since only a fraction of inelastic $p+p$ collisions produce particles which
enter both BBCs, the MB trigger condition biased 
the recorded data sample, so only a
fraction, $f_{\pi^{\circ}}$, of the inclusive $\pi^{\circ}$ yield
was detected.  This fraction was determined from data collected by an additional,
high-$p_T$ trigger which had not been operated
in coincidence with the MB trigger.  This trigger was formed by 
threshold discrimination of the sums of the analog signals from 
overlapping 2$\times$2 groupings of adjacent 2$\times$2 trigger tiles.
As shown in Fig.~\ref{fig:ERT}b, the 
fraction of these high-$p_T$ events with $\pi^{\circ}$'s for which the
MB condition was also satisfied was $0.75\pm0.02$, independent of $p_T$.

In each event,
the two photon invariant mass was calculated for each pairing of
clusters.  Clusters were paired
if the energy 
asymmetry, $|E_1-E_2|/(E_1+E_2)$, was less than 0.8 (PbSc) or 0.7 (PbGl). 
For the PbGl, the pairings were further restricted 
to those clusters 
which were identified as electromagnetic via the shower profile and 
time-of-flight.
The $\pi^{\circ}$ yield was extracted by integrating the invariant mass spectrum
over a region around the $\pi^{\circ}$ mass.  
The background contribution in each $p_T$ bin
was 
estimated and then subtracted by fitting the invariant mass distribution 
outside the peak region (PbSc) or using the mixed event 
technique (PbGl).  For the PbSc and the PbGl, the background to signal ratio 
varied with increasing $p_T$ from 1 to 0.1 and 1 to 0.03, respectively.

The raw yields were corrected for the $p_T$ smearing arising from the EMCal energy 
resolution and the steeply falling spectrum; and for the losses due to 
the disabled towers, the incomplete azimuthal coverage, the energy asymmetry cut,
and the photon identification cut (PbGl).
The correction for these effects, $C_{\rm reco}$, 
was calculated with Monte Carlo simulations
which contained the configuration of the active EMCal towers.
The energy and position of the
decay photons 
were smeared 
with the measured test beam resolutions~\cite{nim_emc} augmented by a constant
energy smearing of 5\% (12\%) for the PbSc (PbGl) to match the response of
the EMCal.

The correction for the losses due to conversions of 
decay photons, $C_{\rm conv}$, was determined to be 4\% (PbSc) and 9\% (PbGl)
by using a {\sc GEANT3}~\cite{geant} simulation of 
the PHENIX detector.  The same simulation, using $p\!+\!p$ events   
from the {\sc PYTHIA} generator~\cite{pythia}, 
showed that the contribution of $\pi^{\circ}$'s from secondary 
interactions was negligible and that the contribution from 
decays of other hadrons ({\it e.g.}, $K^{0}$ and $\eta$ mesons) 
was less than 6\%. 
The $\pi^{\circ}$ spectrum was not corrected for these decays.

The integrated luminosity, $\hat{\cal L}$, was determined from the 
number of MB
events using an absolute calibration of the trigger cross section obtained via the van 
der Meer 
scan technique~\cite{vdmeer}. In a scan, the transverse profile of the
beam overlap is measured by sweeping one beam 
across the other in steps while monitoring the MB trigger rate.  This 
information,
the bunch intensities of the two beams ($\sim$$10^{11}$/bunch), and the 
revolution frequency (78~kHz) are then used 
to compute the instantaneous luminosity. 
The trigger cross section is
the ratio 
of the MB trigger rate 
when the beams were overlapping maximally to the 
instantaneous luminosity.  Based on three scans,
this cross section
was $21.8\pm0.9$~(2.8)~mb 
at the 68.5\%~(95\%) confidence level with an
absolute error of 0.7~mb.
From the linear
sum of the absolute error and half of the 95\% confidence level, point-to-point 
systematic
error, an error of 9.6\% was assigned for the luminosity normalization.

During the run, the maximum and average instantaneous luminosities
were $1.5\! \times \! 10^{30}$ and
$0.5 \! \times \! 10^{30}~{\rm cm}^{-2}\cdot{\rm s}^{-1}$, respectively.  
Contributions
from multiple collisions per bunch crossing and beam-gas interactions 
were negligible.
The MB trigger sample of 16 million events corresponded to 
0.7~${\rm nb}^{-1}$.  
As computed from the fraction of recorded MB events which also 
met the 2$\times$2 high-$p_T$ trigger condition ($\sim$1/47), 
the 18 million high-$p_T$ triggered events corresponded 
to an effective luminosity of 39~${\rm nb}^{-1}$.

The invariant differential cross sections obtained from the MB and 
high-$p_T$ samples were consistent within the statistical errors
over the $p_T$ region of overlap ($p_T$$\le$5.5~GeV/$c$).  Moreover, the 
results determined independently from the PbSc and the PbGl data samples 
were consistent.  The main sources of the point-to-point systematic 
uncertainty in the two measurements are summarized in 
Table~\ref{tab:errors} for a low and a high 
$p_T$ bin.  The total error was computed as the quadrature sum of the 
statistical and point-to-point systematic errors.

\begin{figure}[ht]
\centerline{\epsfig{file=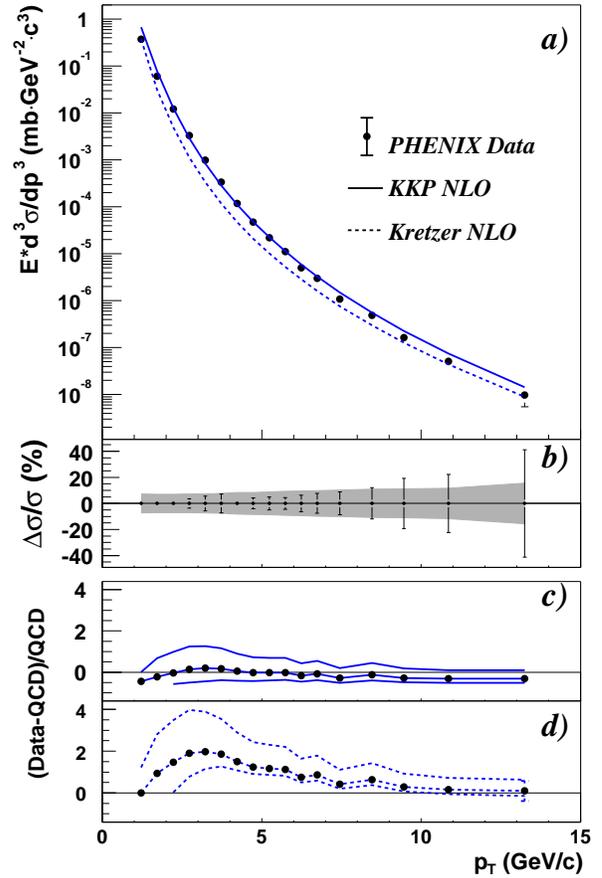,width=1.0\linewidth}}
\caption{a) The invariant differential cross section for inclusive $\pi^{\circ}$ 
production (points) and the results from 
NLO pQCD calculations with equal renormalization and factorization scales of $p_T$ 
using the ``Kniehl-Kramer-P\"{o}tter'' (solid line)  and ``Kretzer'' (dashed line) 
sets of fragmentation functions.  
b) The relative statistical (points) and point-to-point systematic (band) errors.  
c,d)  The relative difference between
the data and the theory using KKP (c) and Kretzer (d) fragmentation functions with
scales of $p_T$/2 (lower curve), $p_T$, and 2$p_T$
(upper curve).  In all figures, 
the normalization error of 9.6\% is not shown.} 
\label{fig:qcd} 
\end{figure}

From the MB and the high-$p_T$ trigger samples for 
$p_T$ below and above 4~GeV/$c$, respectively, 
Table~\ref{tab:cross} tabulates the cross section and the 
errors obtained by averaging the PbSc and PbGl results
using the total error for the weighting.  
Figs.~\ref{fig:qcd}a and~\ref{fig:qcd}b 
show this combined result and 
its fractional statistical and systematic 
uncertainties ($\Delta\sigma/\sigma$).
The data are well parameterized by a power-law 
form \mbox{$A \cdot (1 \! + \! p_T/p_0)^{-n}$} 
with parameters of $A \! \! = \! \! 386$~mb$\cdot$GeV$^{-2}$$\cdot$$c^{3}$, 
$p_0 \! \! = \! \! 1.219$~GeV/$c$, and 
$n \! \! = \! \! 9.99$.

In Fig.~\ref{fig:qcd}, our results are compared with 
NLO pQCD calculations~\cite{acgg,jsv,ddf}.
The basic concept underlying these calculations is the factorization
of the cross section 
into parton distribution functions for the 
protons, parton-to-pion fragmentation functions, and short-distance
partonic hard-scattering cross sections
which can be evaluated
using perturbative QCD. 
Because of this factorization, the calculations depend on unphysical, factorization and
renormalization scales which are of the order of the hard scale $p_T$.
This dependence is reduced as higher order terms 
are included in the perturbation
expansion.
For a calculation truncated at a given order, this dependence serves as a gauge 
for the uncertainty in its results.

The calculations in Fig.~\ref{fig:qcd} have been performed 
with equal
renormalization and factorization scales of $p_T$/2, $p_T$, and 2$p_T$
by using the 
CTEQ6M~\cite{cteq6m} set of parton distribution functions 
and two sets of fragmentation functions.
In general, these calculations are consistent with the data, even at
low $p_T$ ($<$2 GeV/$c$) 
where the theory might be expected to be less applicable.
On closer inspection, as shown Fig.~\ref{fig:qcd}c and~\ref{fig:qcd}d,
the calculation with the ``Kniehl-Kramer-P\"{o}tter'' (KKP)
set of fragmentation functions~\cite{frag_theory:kkp} agrees with 
our data better than the calculation with the
``Kretzer'' set~\cite{frag_theory:kretzer} does.
These two sets differ mainly in that the gluon-to-pion
fragmentation function, $D_g^{\pi}$, is greater in the KKP set.  This
difference is exhibited primarily at low $p_T$ because of the dominance of 
gluon-gluon and gluon-quark
interactions for $p_T$ below $\sim$10~GeV/$c$~\cite{acgg}.
Our measurement thus may impose a meaningful constraint on $D_g^{\pi}$.

%

%

In summary, the invariant differential cross section for inclusive neutral pion production 
in $p\!+\!p$ collisions at $\sqrt{s}\!=\!200$~GeV
was measured at mid-rapidity ($|\eta|\!<\!0.35$) as a function of $p_{T}$ up 
to $\sim$14~GeV/$c$.  These results were compared with two NLO pQCD 
calculations which differed in the choice of fragmentation 
functions.  Over the full range of $p_T$, the calculations were consistent
with the result within the uncertainty of the calculations as 
judged from the scale dependence, although 
the results favored a larger gluon-to-pion fragmentation 
function.

%


We thank the staff of the Collider-Accelerator and Physics
Departments at BNL for their vital contributions.  We thank
Werner Vogelsang and Stefan Kretzer for their interest and 
input.  We acknowledge support from the Department of Energy
and NSF (U.S.A.), MEXT and JSPS (Japan), CNPq and FAPESP
(Brazil), NSFC (China), CNRS-IN2P3 and CEA (France), BMBF, DAAD,
and AvH (Germany), OTKA (Hungary), DAE and DST (India), ISF
(Israel), KRF and CHEP (Korea), RAS, RMAE, and RMS (Russia), VR
and KAW (Sweden), U.S. CRDF for the FSU, US-Hungarian
NSF-OTKA-MTA, and US-Israel BSF.


\def\IJMPA{{Int. J. Mod. Phys.}~{\bf A}}
\def\EPJ{{Eur. Phys. J.}~{\bf C}}
\def\JPG{{J. Phys}~{\bf G}}
\def\JHEP{{J. High Energy Phys.}~}
\def\NCA{Nuovo Cimento~}
\def\NIM{Nucl. Instrum. Methods~}
\def\NIMA{{Nucl. Instrum. Methods in Phys. Research}~{\bf A}}
\def\NPA{{Nucl. Phys.}~{\bf A}}
\def\NPB{{Nucl. Phys.}~{\bf B}}
\def\PLB{{Phys. Lett.}~{\bf B}}
\def\PLC{Phys. Repts.\ }
\def\PRL{Phys. Rev. Lett.\ }
\def\PRD{{Phys. Rev.}~{\bf D}}
\def\PRC{{Phys. Rev.}~{\bf C}}
\def\ZPC{{Z. Phys.}~{\bf C}}

\begin{table}
\caption{
Summary of the sources of systematic errors on the $\pi^{\circ}$ yields
and the total systematic error for $p_T$ of~1.2 and 10.9~GeV/$c$.
The normalization error of 9.6\% is not listed.
}
\begin{tabular}[hb]{lcccc}
                   & \multicolumn{2}{c}{\% Error (PbSc)} & \multicolumn{2}{c}{\% Error (PbGl)} \\  
$p_T$ (in GeV/$c$) &     1.2     &     10.9     &     1.2     &    10.9       \\ \hline
Energy Scale       &      3      &      11      &      6      &     12        \\
Yield Extraction   &      7      &      4       &      5      &      5        \\ 
Yield Correction   &      3      &      6       &      6      &     11        \\ 
Acceptance Stability &     4.5     &      4.5     &      3      &      2        \\ \hline
Total              &      9      &      14      &     10      &     17        \\
\end{tabular}
\label{tab:errors}
\end{table}

\begin{table} 
\caption{The $p_T^*$ (see text for definition), the invariant differential cross section for inclusive 
neutral pion production in $p\!+\!p$ collisions at
$\sqrt{s} \! = \! 200$~GeV, the statistical uncertainty, and the
systematic uncertainty for each $p_T$ bin.
The absolute normalization error of 9.6\% is not included. }
\begin{tabular}[]{ccccc}
             &              & inv. cross & stat. & syst. \\ 
 $p_T$ bin   &  $p_T^{*}$   & section    & error & error \\ 
 (GeV/$c$)   &  (GeV/$c$)   & (mb$\cdot$GeV$^{-2}$$\cdot$$c^{3}$)       &    (\%)     &     (\%)    \\ \hline
  1.0-1.5    &     1.22     &   $3.73\cdot10^{-1}$ &     1.6     &  7.3        \\
  1.5-2.0    &     1.72     &   $6.05\cdot10^{-2}$ &     1.8     &  7.1        \\
  2.0-2.5    &     2.22     &   $1.22\cdot10^{-2}$ &     2.5     &  7.1        \\
  2.5-3.0    &     2.73     &   $3.31\cdot10^{-3}$ &     3.6     &  7.2        \\
  3.0-3.5    &     3.23     &   $9.98\cdot10^{-4}$ &     5.7     &  7.3        \\
  3.5-4.0    &     3.73     &   $3.39\cdot10^{-4}$ &     7.3     &  7.7        \\
  4.0-4.5    &     4.23     &   $1.19\cdot10^{-4}$ &     2.4     &  8.3        \\
  4.5-5.0    &     4.73     &   $4.73\cdot10^{-5}$ &     4.2     &  8.5        \\
  5.0-5.5    &     5.23     &   $2.21\cdot10^{-5}$ &     5.0     &  8.7        \\
  5.5-6.0    &     5.74     &   $1.11\cdot10^{-5}$ &     4.5     &  9.2        \\
  6.0-6.5    &     6.24     &   $5.00\cdot10^{-6}$ &     6.3     &  9.5        \\
  6.5-7.0    &     6.74     &   $3.00\cdot10^{-6}$ &     7.7     &  9.8        \\
  7.0-8.0    &     7.45     &   $1.08\cdot10^{-6}$ &     8.8     &  10.1       \\
  8.0-9.0    &     8.46     &   $4.85\cdot10^{-7}$ &    12.0     &  10.8       \\
  9.0-10.0   &     9.46     &   $1.64\cdot10^{-7}$ &    19.3     &  11.0       \\
 10.0-12.0   &    10.86     &   $5.07\cdot10^{-8}$ &    22.3     &  11.7       \\
 12.0-15.0   &    13.25     &   $9.76\cdot10^{-9}$ &    41.3     &  15.6       \\
\end{tabular}
\label{tab:cross}
\end{table}



\end{multicols}

\end{document}